\documentstyle[epsfig,12pt]{article}
\begin{document}
\begin{center}
{\bf Comment on Fermionic and Bosonic Pair-creation in an External Electric Field 
at Finite Temperature} \\ [16pt]   
Avijit Kumar Ganguly\footnote{avijit@cts.iisc.ernet.in}\\
Centre for Theoretical Studies \\
Indian Institute of Science, Bangalore 560012\\ 
\end{center}
\baselineskip=12pt

\begin{abstract} 
\noindent 
We show that contrary to the claim made by Hallin and Liljenberg in Phys. Rev. \underline{D52}
1150,(1995), (hep-th/9412188) the thermal correction to the {\em thermal decay or pair production rate}
 for a system placed in a heat bath in the presence of an external electric field, is always nonzero in
 the finite as well as infinite time limit. Using the formalism outlined there, we reestimate the 
decay rate for different values of temperature, mass and time.We also try to identify the parameter
 ranges where the quantity of interest agrees with that  computed previously, at high temperature
(in the infinite time limit), from the imaginary part of the effective action using imaginary time and 
real time formalism of thermal field theory. We also point out that in the strictly infinite time limit, 
the correct decay rate as obtained from the work of Hallin et. al.  tends to diverge.
\end{abstract}

\section{Introduction.}
The study of pair-creation phenomena in the presence of an external field in a heat bath has attracted 
some attention in the last few years. There have been attempts to find out the temperature dependence 
of the decay rate by various groups (see for instance references \cite{yildiz} to \cite{gan}).\\

\noindent 
There exists many formalisms to calculate the pair-production rate for such a system. One of them is 
to calculate the effective action, the imaginary part of which is related to the vacuum persistence probability
and the other is to compute the density matrix using functional Schr\"{o}dinger representation and from thence
calculate the production rate. In the past, there had been investigations where the imaginary part of the 
effective action was calculated using the real time formalism or Thermo Field Dynamics (TFD)\cite{loewe} and 
also the imaginary time formalism \cite{gan} of finite temperature field theory. On the other hand J.Hallin and 
Per Liljenberg had derived the vacuum persistence probability for such a system employing the elegant approach
of functional Schr\"{o}dinger representation--in reference \cite{lil}. Unfortunately the results of these 
investigations do not see to agree with each othre. Some of the studies have found a finite thermal correction 
to the production amplitude whereas other studies have claimed to have not found any. In references \cite{loewe}
and ~\cite{gan} a thermal correction to the decay rate have been found and the results of the two almost match
except for a probable typographical error.\\

\noindent 
On the other hand the results of references \cite{yildiz}, \cite{lil} and \cite{elm} are in contradiction 
to \cite{loewe} and \cite{gan}. In this note we shall concentrate on the findings of reference \cite{lil}. The 
authors of reference \cite{lil} claim that at a fixed temperature, in the infinite time limit (as is assumed in 
the effective action formalism), the thermal correction to pair production rate vanishes thereby contradicting 
the earlier results found using the finite temperature effective action formalism.\\

\noindent 
In this note, starting from eqn.~[208] of reference \cite{lil} --- obtained by the functional Schr\"{o}dinger 
representation (for the fermionic case), we will show that there is a {\em finite correction} to the pair 
production rate, even in the very large time limit at any nonzero temperature. This is in direct contradiction
to the results obtained in \cite{lil}. It must be noted here that though under some approximations this result 
reduces to the results obtained in reference \cite{loewe} and \cite{gan} and we shall discuss the implications of
that shortly. \\

\noindent 
To be little more precise, we find that in the limit of very small time and high temperature the results of 
reference \cite{lil} matches with those of \cite{loewe} and \cite{gan} found in the high temperature limit. On 
the other hand in the studies of \cite{loewe} and \cite{gan} the production rate is calculated in the infinite 
time limit as performed in the strict thermodynamic sense. In this paper we will also point out that the 
production rate when properly computed, turns out to be a function of time and it tends to diverge once the 
infinite time limit is assumed.

\section{Calculation}                                                                                 
In this section we start from the expression of probability $P(0,0;t_f)$, of finding the system with no fermion 
anti-fermion pairs in a state after a time $t= t_f$, as is given in eqn.~[208] of Hallin and Liljenberg \cite{lil}, 
\begin{equation}                                                                             
P(0,0;t_f) = exp \left[ C({\bar \beta})  + \frac{2 \bar{V}}{2 (2 \pi^2)}  
\int_{\bar{m}^2 }^\infty d \Lambda \int_{-\bar{t_f}^2 }^0 
d {\bar{p}}^3  ln \frac{1+ th (2 - 4 e^{- \pi \Lambda} )}{ (1 + th )^2} \right]
\label{p}
\end{equation}
                                                                        
\noindent 
For the benefit of the readers following reference \cite{lil} we re-introduce the notations as given 
there in \cite{lil} 
\begin{eqnarray}                                                                             
th = tanh \left[ \frac{\bar{\beta} \bar{ \omega }_{0}^{g} }{2}\right], 
\hspace{20pt}
\bar{ \omega }_{0}^{g} = \sqrt{ \Lambda + (\bar{p}^3 + \bar{t_f} )^2 } 
\nonumber \\ 
\Lambda = (\bar{p^1})^2 + (\bar{p^2})^2 + (\bar{p^3})^2 + (\bar{m})^2 
, ~~~~\bar{V} = V (eE)^{3/2},~~~~~~~~
 \bar{p} = \frac{p}{ \sqrt{eE}} 
\\ 
\bar{m}=\frac{m}{\sqrt{eE}},
\hspace{30 pt} \bar{t_f} = t_{f} (eE)^{1/2},\hspace{30 pt} \bar{\beta}= \beta
\sqrt{eE}.
\label{defn}  \nonumber 
\end{eqnarray}                                                             
With these definitions eqn.[\ref{p}] can easily be cast in the following form
\begin{eqnarray}
P(0,0;t_f) = exp \left[ C({\bar \beta})  + \frac{ \bar{V}}{ (2 \pi)^2
}   \int_{\bar{m}^2 }^\infty d \Lambda \int_{-\bar{t_f}^2 }^0 
d {\bar{p}}^3  ln\left[1 - \frac{4 th }{ (1 + th )^2}e^{- \pi \Lambda} \right]
\right].                           
\label{P1}                  
\end{eqnarray}                                                                         
Since the quantity inside the logarithm in eqn.[\ref{P1}] is less than one we can expand the logarithm to 
arrive at
\begin{eqnarray}
P(0,0;t_f) = exp \left[ C({\bar \beta})  - \frac{ \bar{V}}{ (2 \pi)^2}
   \int_{\bar{m}^2 }^\infty d \Lambda e^{- \pi \Lambda} \int_{-\bar{t_f}^2 }^0 
d {\bar{p}}^3  \left[1 - e^{2 \bar{\beta} {\omega_{0}}^g  } \right]
\right].                           
\label{P2}
\end{eqnarray}
The expression for vacuum persistence probability, in the Schwinger sense \cite{sch} is obtained from the 
eqn.[\ref{p}] as
\begin{eqnarray}
W = \lim_{t_f \rightarrow \infty}  \frac{ln  P (0,0,t_f) } {V t_{f}}.
\label{s1}
\end{eqnarray}
\\
\noindent 
In \cite{lil} the infinite time limit was taken inside the integrand before carrying out the 
integration to find out the explicit $t_f$ dependence of the resulting expression. Though in some cases this 
procedure might give identical result with the case where the limit is taken afterwards, here it does not seem 
to be so. Therefore, in order to compute the production rate W, unlike the authors of \cite{lil}, we shall 
evaluate the integral retaining the explicit $t_{f}$ dependence before dividing it by $t_f$ and finally taking 
the limit $t_f \rightarrow \infty$. In order to simplify the situation a bit, from now on, we will work for a 
while with  the integral inside eqn.[\ref{P2}]. It is given by,
\begin{eqnarray}
\int_{\bar{m}^2 }^\infty d \Lambda e^{- \pi \Lambda} \int_{-\bar{t_f}^2 }^0 
d {\bar{p}}^3  \left[1 - e^{2 \bar{\beta} {\omega_{0}}^g  } \right] =
\int_{\bar{m}^2 }^\infty d \Lambda e^{- \pi \Lambda} \int_{-\bar{t_f}^2 }^0 
d {\bar{p}}^3   -
\int_{\bar{m}^2 }^\infty d \Lambda e^{- \pi \Lambda} \int_{-\bar{t_f}^2 }^0 
d {\bar{p}}^3 \,\, \nonumber  \\  \left[e^{2 \bar{\beta} {\omega_{0}}^g  } 
\right].
\label{P3}
\end{eqnarray}

\noindent 
The first integral here gives the zero temperature Schwinger pair production probability and since the 
controversy in the literature is about the finite temperature piece we will concentrate only on the second term 
in eqn.[\ref{P3}]. In the second integral the quantity $\bar{{\omega_{0}}^{g}}$ is given by 
${\bar{\omega }}_{0}^{g} = \sqrt{ \Lambda + (\bar{p}^3 + \bar{t_f} )^2 }$ and it also appears in the exponent 
thus making the evaluation with exact $t_f$ dependence more difficult. Therefore to circumvent this difficulty we 
propose to rewrite the exponential using an integral transform of the following form 
\cite{dinge} \footnote{This follows from the definition of the Bessel function.}
\begin{eqnarray}
e^{- \alpha \sqrt{s}} = \frac{\alpha}{2 \sqrt{\pi}} \int_{0}^\infty 
e^{ - us - \frac{\alpha^2}{4u}}  \frac{d u}{u^{\frac{3}{2}}}. 
\end{eqnarray} 
Now if we identify $2{\bar{\beta}}$ with $\alpha$ and the quantity ${\bar{\omega_{0}}}^{g}$ as $\sqrt{s}$, we 
can write
\begin{eqnarray}
e^{- {\bar{\beta}} \bar{{\omega_{0}}^{g}}} = \frac{\bar{\beta}}{2 \sqrt{\pi}} 
\int_{0}^\infty 
e^{ - u \left[ \Lambda + (\bar{p}^3 + \bar{t_f} )^2 \right] 
- \frac{{\bar{\beta}}^2}{4u}}  \frac{d u}{u^{\frac{3}{2}}}.
\label{P4}
\end{eqnarray} 

\noindent 
With the aid of eqn.[\ref{P4}], we obtain from the second integral (eqn. [\ref{P3}]) upon performing
the $\Lambda$ integration
\begin{eqnarray}
I &=& \frac{\bar{\beta}}{ \sqrt{\pi}} 
 \int_{-\bar{t_f}}^{0} d {\bar{p}}^3
\int_{0}^{\infty} \frac{d u}{u^{\frac{3}{2}}} \int_{\bar{m^2}}^\infty  d \Lambda 
e^{-\Lambda ({\pi +u}) }
e^{ - u \left[(\bar{p}^3 + \bar{t_f} )^2 \right] 
- \frac{{\bar{\beta}}^2}{4u}} \nonumber \\                                                                     
&=& \frac{\bar{\beta}}{ \sqrt{\pi}} e^{ - {\bar{m}}^2 \pi } \int_{-\bar{t_f}}^{0} 
d {\bar{p}}^3 
\int_{0}^{\infty} \frac{d u}{u^{\frac{3}{2}}}
\frac {e^{ - u \left[ {\bar{m}}^2 + (\bar{p}^3 + \bar{t_f} )^2 \right]- 
\frac{{\bar{\beta}}^2}{4u}} } {\left( \pi + u \right) }.
\label{P5}
\end{eqnarray} 

\noindent 
Hereafter one can interchange the order of the integrations and carry out the ${\bar{p}}^{3}$
integration first. Upon performing this integration the resulting expression can be cast in the form of 
an error function with $\sqrt{u} \times \bar{t}_f $ as its argument. Rewriting the error function as a 
series we obtain 
\begin{equation}
I= \frac{\bar{\beta}}{\sqrt{\pi}} e^{- \bar{m}^2 \pi}
\sum_{k=1}^{\infty}
 \frac{\left( \bar{t}_f \right)^{2k-1} \left( -1\right)^{k+1}}{\left( 2k-1\right)\left(k 
-1\right)!}
\int_{0}^{\infty} \frac{du u^{(2k-5)/2}}{u+ \pi}
e^{- u \bar{m}^2 - \frac{\bar{\beta}^2}{4u}} 
\label{SP5}
\end{equation}

\noindent 
In order to write the pair production rate in a more compact form, we make the following change of 
variables $u \rightarrow z (\frac{\bar{\beta}}{2})^{2}$ in eqn.[\ref{SP5}] and use the definition of pair 
production rate as given in eqn.[\ref{s1}] without taking the limit $t_f \rightarrow \infty$ before hand. 
And the resulting expression is, 
\begin{equation}
W= \frac{eE}{\beta^2 (2 \pi)^2 }
\frac{2}{\sqrt{\pi}} e^{- \bar{m}^2 \pi}
\sum_{k=1}^{\infty}
 \frac{\left( \bar{t}_f \bar{\beta} \right)^{2k-2} \left( -1\right)^{k+1}}{\left( 
2k-1\right)\left(k -1\right)!}
\int_{0}^{\infty} \frac{dz z^{(2k-5)/2}}{ \frac{z}{4}+ \frac{\pi}{\bar{\beta}^2}}
e^{- z ((\bar{\beta}\bar{m})/2)^2 - \frac{1}{z}} .
\label{SP6}
\end{equation}
It can be seen from eqn.[\ref{SP6}] that for any nonzero value of $t_f$ the pair production rate, contrary to the 
claim made in \cite{lil}, is always nonzero. At zero temperature, i.e, when $\beta = \infty$, eqn.[\ref{SP6}] goes 
to zero because of the $(\beta)^2$ sitting in the exponent. And it conforms to our general expectation that as 
temperature tends to zero the temperature dependent correction to the rate should also vanish. On the other hand 
at a nonzero temperature, in the thermodynamic sense, as $t_f \rightarrow \infty$ the rate becomes divergent. This 
is an unique feature that comes out of the treatment of reference \cite{lil}. This is the main result of this 
paper. In the remaining portion of the paper we shall try to evaluate this expression numerically for different 
values of the parameters and write down the approximate analytic forms for each of these cases. \\

\section{Discussion and Results}

\noindent 
In this section we evaluate $W$ for three different values of the parameters namely 
(a) $t_f \ll 1 ,~ \bar{m} \ll 1$ and $\bar{\beta} \sim  O(1)$ or little more, 
(b) $\bar{t}_f \gg 1$ but $\bar{\beta} \bar{t_f}   \ll 1 $ and $\bar{m} \ll 1$ and lastly 
(c) both $\bar{\beta}$ and $\bar{m} $ are $ \ll 1$ but $\bar{t}_f \gg 1$. In the last case we would vary 
$\bar{t}_f$ to show how $W$ changes with its variation.\\

\noindent
We shall start with case (a) and then move over to the other two cases. In order to derive an approximate 
analytical form for $W$ for this case we start from eqn.[\ref{P5}] instead of eqn.[\ref{SP6}] as given in this 
text. Since in this case both $t_f$ and $\bar{m}$ are assumed to be $\ll 1$ and the parameter $\beta$ is of the 
order of one or slightly more, the quantity 
$\bar{\beta} \sqrt{ \left[ {\bar{m}}^2 + (\bar{p}^3 + \bar{t_f} )^2 \right] }$ is very small. Since the integral 
there varies from $0$ to $\infty$ we would like to approximate the quantity $(u + \pi)$ in the denominator by u 
to arrive at
\begin{eqnarray}
\frac{\bar{\beta}}{ \sqrt{\pi}} e^{ - {\bar{m}}^2 \pi } \int_{-\bar{t_f}}^{0} d 
{\bar{p}}^3 
\int_{0}^{\infty} \frac{d u}{u^{\frac{5}{2}}}
e^{ - u \left[ {\bar{m}}^2 + (\bar{p}^3 + \bar{t_f} )^2 \right]- 
\frac{{\bar{\beta}}^2}{4u}}  .
\label{P6}
\end{eqnarray}

\begin{figure}[h]
\begin{center}{\mbox{\epsfig{file=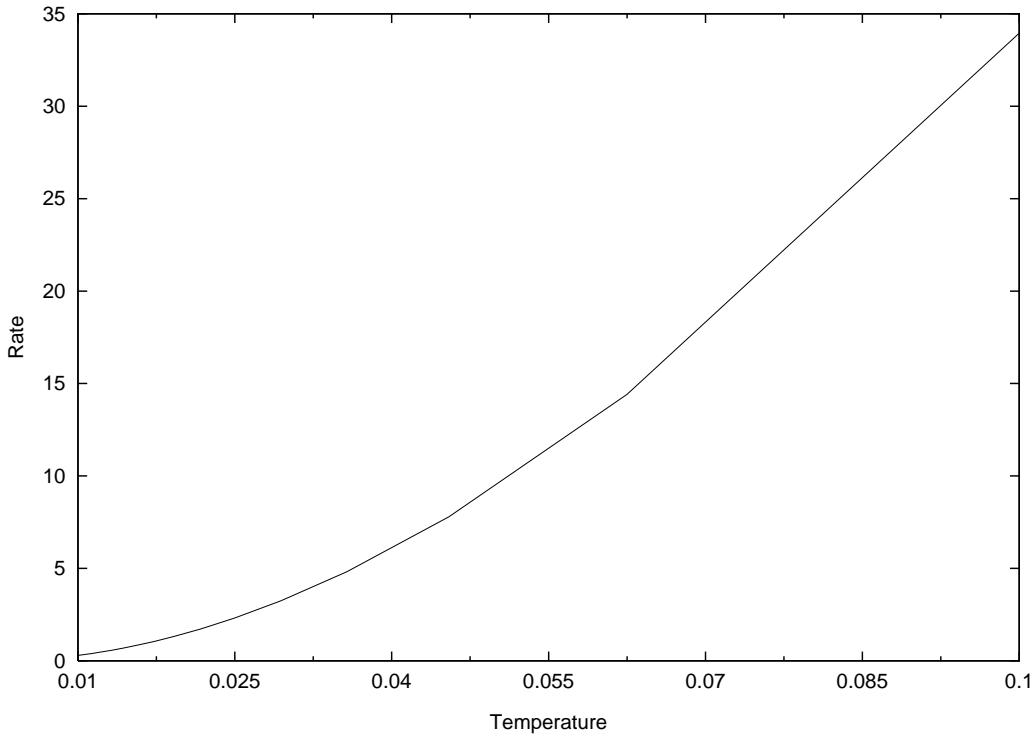,width=400pt}}}\end{center}
\caption{ Production rate vs temperature for $t_f = 1.0 \times 10^{-3}$ and $\bar{m} = 1.0 \times 10^{-2}$.}
\label{parabola}
\end{figure}

\noindent  
Now we can perform the $u$ integration by using the formula for modified Bessel-function,
$\int_{0}^{\infty}  u^{ \nu -1}
e^{ - u \gamma - \frac{\delta}{u}}~du = 2 \left[ \frac{\delta}{\gamma} 
\right]^{\frac{\nu}{2}} K_{\nu} \left[2 \sqrt{\delta \gamma} \right]. $
And using this form of the modified Bessel function eqn.[\ref{P6}] now comes out to be 
\begin{eqnarray} 
\frac{\bar{\beta}}{ \sqrt{\pi}}
 e^{ - {\bar{m}}^2 \pi } 
 \int_{-\bar{t_f}}^{0} d {\bar{p}}^3 ~~ 
2 \left( \frac{\bar{\beta}^2}{\left[ 
{\bar{m}}^2 + (\bar{p}^3 + \bar{t_f} )^2 \right]} \right)^{- \frac{3}{4}}
K_{ - \frac{3}{2}} \left[ 2 \bar{\beta}
 \sqrt{ \left[ {\bar{m}}^2 + (\bar{p}^3 + \bar{t_f} )^2 \right]} \right]  .
\label{P7}
\end{eqnarray}\\
For the range of parameters that we are interested in the argument of the modified Bessel-function tends to be very 
small. In this limit one can replace the modified Bessel-function by its approximate form, 
$K_{\frac{3}{2}}(z) \approx  \frac{1}{2} \frac{\Gamma\left(\frac{3}{2} \right)}{(\frac{z}{2})^{\frac{3}{2}}}$. 
Using this relation eqn.[\ref{P5}] becomes
\begin{eqnarray} 
I &=& \frac{\bar{\beta}}{ \sqrt{\pi}}
 e^{ - {\bar{m}}^2 \pi } 
 \int_{-\bar{t_f}}^{0} d {\bar{p}}^3 ~~ 
2 \left( \frac{\bar{\beta}^2}{\left[ 
{\bar{m}}^2 + (\bar{p}^3 + \bar{t_f} )^2 \right]} \right)^{- \frac{3}{4}}
K_{ - \frac{3}{2}} \left[ 2 \bar{\beta}
 \sqrt{ \left[ {\bar{m}}^2 + (\bar{p}^3 + \bar{t_f} )^2 \right]} \right]   
\nonumber \\ 
 &=& \frac{ \bar{t_{f}} e^{ - {\bar{m}}^2 \pi }}{2 \bar{\beta}^2}
\label{P8}
\end{eqnarray}
For $\bar{m}\ll 1$ the exponential can be approximated by $\approx$ 1. Hence the pair production rate in this 
limit, (using eqn.[\ref{s1}]) turns out to be,
\begin{eqnarray} 
W\approx \frac{eE T^2}{8 \pi^2}.
\label{app1}
\end{eqnarray}
We have also evaluated, numerically, the pair-production rate from eqn.[\ref{SP6}] for 
$\bar{t}_f = 10^{-3}, ~with~ \bar{m} = 10^{-2}$ and have plotted it in figure [\ref{parabola}].
It shows the behavior of W as one moves from low temperature to high temperature. It can be seen from the curve 
that the general trend qualitatively agrees with that given by eqn.[\ref{app1}]. As has already been mentioned 
in the beginning, apart from the numerical factors, this result agrees with that of reference [\cite{loewe}] 
and [\cite{gan}] obtained in the high temperature and infinite time limit from the imaginary part of the thermal 
effective action.\\

\begin{figure}[h]
\begin{center}{\mbox{\epsfig{file=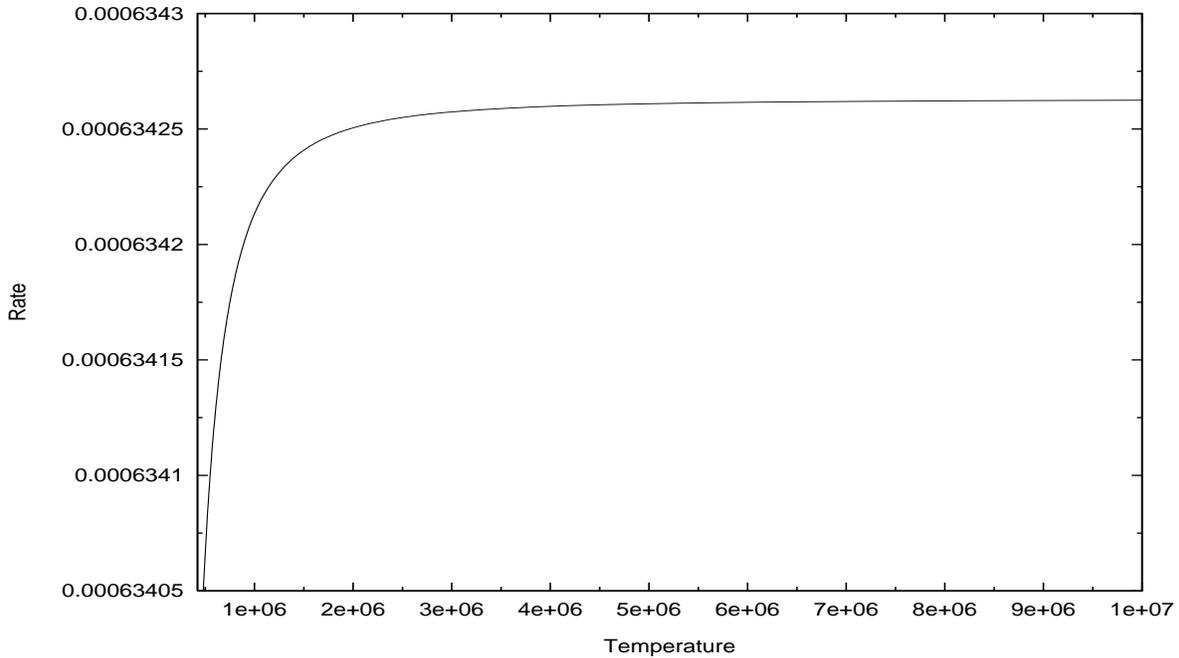,width=300pt}}}\end{center}
\caption{Production rate vs temperature for $t_f=1.0 \times 10^{3} $, $\bar{m}=1.d \times 10^{-6}$}
\label{EXPR}
\end{figure}

\noindent
Having discussed the case for $\bar{t}_f \ll 1$ and $ \bar{m} \ll 1$, we will try to evaluate the pair production 
rate for $\bar{t}_f \gg 1$ but $\bar{\beta} \bar{t}_f \ll 1$ and $\bar{m} \ll 1$. In order to do that, it will be 
convenient if we proceed from eqn.[\ref{SP6}]. The first important thing to notice here is, as 
$\bar{\beta} \bar{t_f} \ll 1$, the dominant contribution to the quantity of our interest comes from the first term
in the summation. Hence, for the dominant contribution, it is sufficient to retain the $k=1$ term where $W$ can be 
approximated by 
\begin{eqnarray} 
W \approx \frac {2 (eE)}{(2 \pi)^2 \sqrt{\pi} \beta^2} e^{- \bar{m}^2 \pi}
\left( \int_{0}^{\infty}
 \frac { z^{- \frac{3}{2} } dz }{ \frac{z}{4} + \frac{\pi}{\beta^2}}   
e^{ -\frac{(\bar{m} \bar{\beta})^2}{4} z -\frac {1}{z}} \right).
\label{app2}
\end{eqnarray}\\
Since the dominant contribution to the exponential comes from $z \approx (\frac{2}{\bar{\beta}\bar{m}})$, 
which for the parameter range of our interest turns out to be quite large, we will proceed as before by 
approximating the denominator by $z$ and then carrying out the $z$ integration using the modified Bessel function 
of fractional order. On using the exact form of the Bessel function eqn.[\ref{app2}] approximately comes out to be 
\begin{eqnarray} 
W \approx \frac {(eE)}{(2 \pi)^2 \beta^2} e^{- \bar{m}^2 \pi}
e^{ -(\bar{m} \bar{\beta})}.
\label{app3}
\end{eqnarray}
We have estimated $W$ numerically for $\bar{m} = 10^{-6}$ and $\bar{t}_f = 10^{-3}$ for various values of 
temperature ranging from $10^6$ to $10^7$ and have plotted them in figure [\ref{EXPR}]. It can be seen that the 
shape of the curve is basically due to the function $e^{ -(\bar{m} \bar{\beta})}$ 
as is given in eqn.[\ref{app3}]. \\

\begin{figure}[h]
\begin{center}{\mbox{\epsfig{file=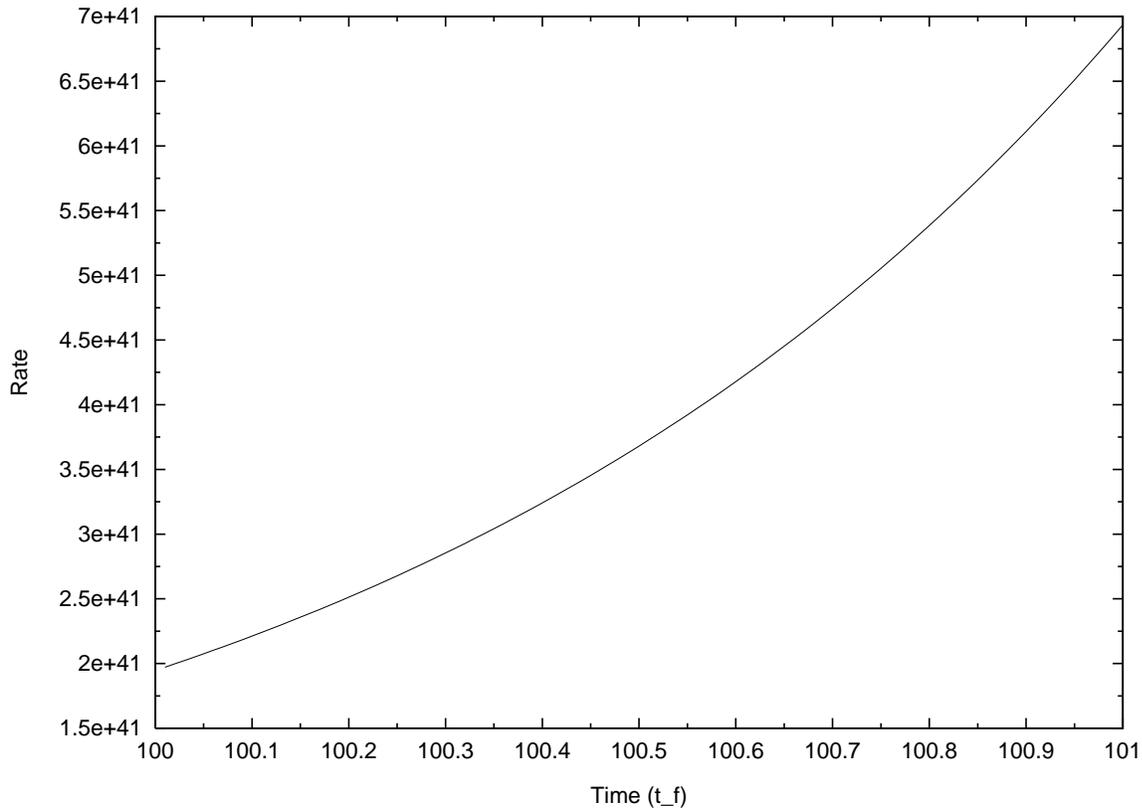,width=400pt}}}\end{center}
\caption{Production rate vs $t_f$ for  Temperature= $1.0 \times 10^{-4}$ , $\bar{m}=1.d \times 10^{-6}$ }
\label{time}
\end{figure}

\noindent
Lastly in figure [\ref{time}] we have plotted $W$ for $\bar{\beta}=10^{-4}$ and $\bar{m}=10^{-6}$
as a function of $t_f$, ranging from 100 to 101 in arbitrary units. One can see from this curve that as time 
$t_f$ increases $W$ also keeps on increasing. So in the strict thermodynamic limit as $t_f \rightarrow \infty$ 
the $W$ would become divergent.  

{\section{Conclusion}}

\noindent
Starting from eqn.[208] of ref.\cite{lil} we have shown that for a fermionic system, contrary to the claim
made there (i.e. in \cite{lil}), at any non-zero temperature there is a finite correction to Schwinger's pair 
production rate for any value of $t_f$.It is shown that the final expression (i.e. eqn.[\ref{SP6}]) comes as
a power series in $\bar{\beta} \bar{t}_f$. Though for small values of $\bar{\beta}$ and $\bar{t}_f$, this
expression seems to be finite, but  in the  proper thermodynamic sense it is not.
Since for realistic situations $\bar{\beta}$ can never be zero, on the otherhand  $\bar{t}_f$ in principle 
can approach infinity, hence in the infinite time limit, the quantity $W$ tends to diverge.
The reason behind this apparently contradictory results seems to depend on how one takes the limit $t_f 
\rightarrow \infty$.In our view one should first carry out all the integrations and write down the final result
as a function of $t_f$, before taking the limit $t_f \rightarrow \infty$. Though we have carried out this 
detailed exercise for fermions only but in principle this can be extended for Bosons too.\\

\noindent
We would like to mention here that the same quantity obtained using the finite temperature effective action
formalism ( where infinite time limit is implicitly assumed) using both Thermo Field Dynamics and imaginary
time formalisms ( see e.g. \cite{loewe} and \cite{gan}) yields a nonzero finite result. In view of the previous
results we find this corrected result of reference \cite{lil}to be quite amazing and in our view this deserves
further study. Particularly one should try to understand why the different formalisms are giving different answers
and which one of them is correct. But in case the corrected result of reference \cite{lil} which  is obtained through the 
elegant formalism of Functional Schr\"{o}dinger Representation turns out to be right (though we have some reservatio
ns about it), it can certainly be used to estimate the thermal contribution to the soft processes dominated production mechanisms in the pre-equilibrium 
Quark Gluon Plasma production phase of relativistic heavy ion collisions, at the SPS and RHIC.\\

\section{Acknowledgment}
Its a pleasure to thank Sanjay Jain  and Apoorva Patel for many interesting discussions and encouragements
and very useful suggestions made during the course of this work. The Author
would also like to thank S. Konar, for the help rendered during the
preparation of this manuscript.

\end{document}